\documentclass[twocolumn,showpacs,preprintnumbers,amsmath,amssymb,aps,prd,nofootinbib]{revtex4}
\usepackage{graphicx}

\newcommand{\Lie}[0]{{\cal L}\, }
\newcommand{\grad}[0]{\nabla\!}

\newcommand{\ImPt}[1]{{\mathrm{Im}} \left[ #1 \right]}

\begin{document}

\title{Introduction to Isolated Horizons in Numerical Relativity}
\date{\today}
    \author{Olaf Dreyer}
    \email{odreyer@perimeterinstitute.ca}
    \affiliation{Perimeter Institute for Theoretical Physics, 35 King
    Street North, Waterloo, Ontario N2J 2W9, Canada}
    \altaffiliation{Center for Gravitational Physics and Geometry
    and Center for Gravitational Wave Physics, Department of
    Physics, Penn State University, University Park, PA 16802,
    USA}
    \author{Badri Krishnan}
    \thanks{Present address: Max Planck Institut f\"ur Gravitationsphysik, Am
    M\"uhlenberg 1, D-14476 Golm, Germany}
    \email{badkri@aei-potsdam.mpg.de}
    \affiliation{Center for Gravitational Physics and Geometry
    and Center for Gravitational Wave Physics, Department of
    Physics, Penn State University, University Park, PA 16802,
    USA}
    \author{Erik Schnetter}
    \email{schnetter@uni-tuebingen.de}
    \affiliation{Theoretische Astrophysik, Universit\"at T\"ubingen, Auf der
    Morgenstelle, 72076 T\"ubingen, Germany}
    \altaffiliation{Center for Gravitational Physics and Geometry
    and Center for Gravitational Wave Physics, Department of
    Physics, Penn State University, University Park, PA 16802,
    USA}
    \author{Deirdre Shoemaker}
    \thanks{Present address: Department of Physics, Cornell
    University, Ithaca, NY 14853, USA}
    \email{deirdre@astro.cornell.edu}
    \affiliation{Center for Gravitational Physics and Geometry
    and Center for Gravitational Wave Physics, Department of
    Physics, Penn State University, University Park, PA 16802,
    USA}

\begin{abstract}
We present a coordinate-independent method for extracting mass
($M_\Delta$) and angular momentum ($J_\Delta$) of a black hole in
numerical simulations.  This method, based on the isolated horizon
framework, is applicable both at late times when the black hole has
reached equilibrium, and at early times when the black holes are widely 
separated.  Assuming that the spatial hypersurfaces used in a given
numerical simulation are such that apparent horizons exist and have
been located on these hypersurfaces, we show how $J_\Delta$ and 
$M_\Delta$ can be determined in terms of only 
those quantities which are intrinsic to the apparent horizon.  
We also present a numerical method for finding the rotational symmetry 
vector field (required to calculate $J_\Delta$) on the horizon.
\end{abstract}


\maketitle

\section{Introduction}
\label{sec:intro}

Given a numerical simulation of a spacetime containing a black hole,
one is faced with two immediate questions: Where is the black hole,
and what are the values of its parameters, such as mass and angular
momentum?

In analytical considerations, a black hole is often defined via
the event horizon, i.e.\ the future boundary of the causal past of
future null infinity.  While this notion is mathematically elegant
and has led to powerful results in black hole physics, generally
it is not directly useful in numerical evolutions of black hole
spacetimes. The reason is the teleological nature of the event
horizon: it can be constructed only after we have knowledge of the
entire spacetime and spacetime is the \emph{end product} of these
simulations. Furthermore, due to practical limitations, generally
one cannot evolve all the way to future null infinity. For such
reasons, in numerical evolutions, it is more common to use
\emph{apparent horizons} to characterize a black hole. Apparent
horizons are closed two-surfaces on a spatial slice of the
spacetime and are, therefore, well suited to numerical simulations
that evolve data from one spatial slice to another. While it is
possible that there exist no apparent horizons on the spatial slices 
used in a given numerical simulation of a black hole
spacetime\footnote{For example, it has been shown in \cite{wi} that
there exist Cauchy surfaces containing no apparent horizons even in 
the Schwarzschild spacetime}, in practice, this issue does not arise
for most numerical black hole simulations that we are aware of.  
We shall therefore disregard this issue for the rest of this paper
and only deal with partial Cauchy surfaces (in black hole spacetimes) for
which apparent horizons do exist and have been located
\cite{t,bcsst,aclmss,hcm,a,dhm,es}.  A natural
question we want to ask is: can one unambiguously associate black hole
parameters to apparent horizons?

One way to attribute a mass and an angular momentum to a black
hole is to calculate the corresponding ADM quantities at infinity.
The main difficulty is that the ADM mass and angular momentum
refer to the whole spacetime.  In a dynamical situation, such a
spacetime will contain gravitational radiation and it is not clear
how much of the mass or angular momentum should be attributed to
the black hole itself and, if there is more than one black hole,
to each individual black hole.

It is desirable to have a framework that combines the properties
of apparent horizons with the powerful tools available at
infinity.  In the regime when the black hole is isolated in an
otherwise dynamical spacetime, such a framework now exists in the
form of isolated horizons \cite{abf,letter, afk, abl1,abl2}.  In
numerical simulations of, say, black hole collisions, the black
hole would be isolated at early times when the black holes are
well separated, or at late times when the final black hole has
settled down, but radiation is still present in the spacetime.
Isolated horizons provide a way to identify a black hole
quasi-locally, and allow for the calculation of mass and angular
momentum.  It has been shown recently that the formulae for angular
momentum and mass arising from the isolated horizon formalism are
valid even in dynamical situations \cite{ak}. Therefore, the numerical methods
presented in this article are relevant even for fully dynamical black holes.

The aim of this paper is to introduce the isolated horizon
framework in a way that is directly useful in numerical
relativity. In particular, we show how to find isolated horizons
numerically, and how to implement the isolated horizon formulae
for $J_\Delta$ and $M_\Delta$.  A key result of this paper is
that, to find the angular momentum of an isolated horizon, one can
just use the ADM formula for angular momentum but now applied at
the apparent horizon. This is not an assumption, but a rigorous
result obtained by calculating the Hamiltonian generating
diffeomorphisms which reduce to rotational symmetries on the
isolated horizon.  This is completely analogous to what is done at
infinity to obtain the ADM formulae for mass and angular momentum
for asymptotically flat spacetimes.  Indeed, the Hamiltonian
analysis of isolated horizons is an extension of the ADM formalism
to the case where the region of spacetime under consideration has
an inner boundary in the form of an isolated horizon.  The
isolated horizon results for $J_\Delta$ and $M_\Delta$ are also
convenient for practical reasons because their expressions only
involve data defined on the apparent horizon. An important
ingredient in the formula for $J_\Delta$ is an axial symmetry
vector on the apparent horizon.  Therefore we also present and
implement a numerical method for locating Killing vectors on the
horizon.

This paper is organized as follows.  In section \ref{sec:ih} we first
review the notion of isolated horizons and explain how they can be
located in numerical simulations of black hole spacetimes.  We also
explain their relation to apparent horizons and give the formulae for
their mass and angular momentum.  These formulae require that we find
a Killing vector on the apparent horizon.  A method for doing this is
described in section \ref{sec:Killingvec}.  Section \ref{sec:numeric}
describes the practical numerical implementation of these formulae,
and we verify that our method gives the expected results in the simple
case of a boosted Kerr black hole.  Section \ref{sec:discussion}
compares the isolated horizon formulae with other approaches used to
compute mass and angular momentum, and finally in section
\ref{sec:conclusion} we give some possible applications and future
directions.

\section{Isolated Horizons}
\label{sec:ih}

Let us first fix our notation.  The spacetime metric is taken to
have signature $(-,+,+,+)$, and we use geometrical units where $G$
and $c$ are equal to unity.  We will usually use the abstract
index notation, but occasionally, especially for differential
forms, the index free notation will also be used.  The Riemann
tensor ${R_{abc}}^d$ will be defined by the equation
$2\grad_{[a}\grad_{b]}\alpha_c={R_{abc}}^d\alpha_d$ where
$\alpha_a$ is an arbitrary co-vector.  We work on a spacetime that
is foliated by a spatial three manifold $\Sigma$ which is a partial
Cauchy surface.  The spacetime
is thus of the form $\Sigma\times \mathbb{R}$.  The three-metric and extrinsic
curvature on a spatial slice $\Sigma$ are respectively denoted by
$\gamma_{ab}$ and $K_{ab}:=-{\gamma_a}^c{\gamma_b}^d\grad_c T_d$
where $T^a$ is the unit timelike normal to $\Sigma$.  The
two-metric on the apparent horizon is called $q_{ab}$.  All
manifolds and fields will be taken to be smooth.  Finally, though
it is quite easy to include matter fields, we will be interested
in vacuum spacetimes only.  This paper is reasonably self
contained, however,  for precise and detailed definitions and
proofs we refer the reader to \cite{abf,afk,abl1,abl2}.  In this
paper we shall focus on the basic physical ideas underlying the
isolated horizon framework and its application to numerical
relativity.

\subsection{Motivation, definition and basic results}
\label{sec:definitions}

For completeness and to fix notation, let us start by reviewing
the definition of apparent horizons.  Let $T^a$ be the unit
timelike vector field orthogonal to a partial Cauchy surface  $\Sigma$.
Given a closed two-surface $S\subset\Sigma$, we have the unique
unit outward-pointing spacelike normal $R_a$ which is tangent to
$\Sigma$.  We are assuming here that one can define what is meant by
the inside and outside of $S$.  This can be done, for instance, by
imposing certain global conditions on the spacetime and the partial Cauchy
surfaces under consideration.  These conditions include: asymptotic
flatness,  strong future asymptotic predictability of the spacetime
with the Cauchy surfaces under consideration and certain topological
restrictions on $\Sigma$. Since the focus of this paper is not on
such global topological properties of spacetimes, we shall not spell out these
assumptions precisely and instead refer the reader to \cite{hawk}. 
It is worth mentioning that the inside and outside $S$ can also be 
defined quasi-locally without using global 
assumptions such as the ones mentioned above.  This has been done by 
Hayward \cite{sh} using 
the notion of trapping horizons which we discuss in a later section.

\begin{figure}
  \begin{center}
  \includegraphics[height=6cm]{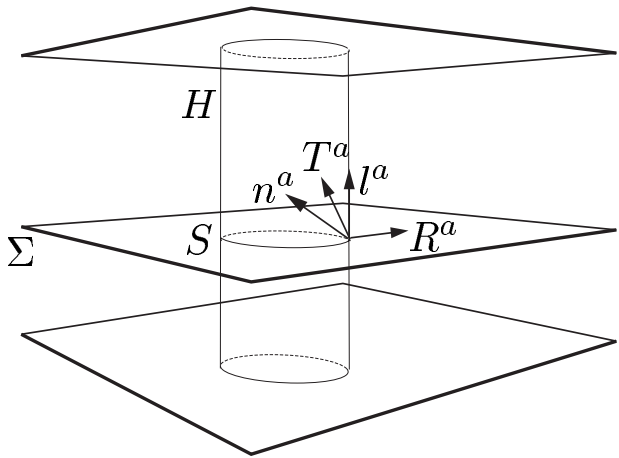}
  \caption{\small{The figure shows an apparent horizon $S$ embedded
  in a spatial slice $\Sigma$.  $T^a$ is the unit timelike normal to
  $\Sigma$ and $R^a$ is the outward pointing unit spatial normal to $S$
  in $\Sigma$; $\ell^a$ and $n^a$ are the outgoing and ingoing null
  vectors, respectively.  The vector $m^a$ (not shown in the figure) is tangent
  to $S$.  $H$ is the world tube of apparent horizons.}}\label{fig:ihah}
  \end{center}
\end{figure}

Let $q_{ab}$ be the induced Riemannian two-metric on $S$
and $\epsilon_{ab}$ the area two-form on $S$ constructed from
$q_{ab}$.  We can construct a convenient basis for performing
calculations at points of $S$ in a natural way (see figure
\ref{fig:ihah}).  First, define the outgoing and ingoing null
vectors
\begin{equation}\label{eq:normals}
    \ell^a := \frac{1}{\sqrt{2}}(T^a + R^a)\quad
    \textrm{and} \quad n^a:=\frac{1}{\sqrt{2}}(T^a - R^a) \, .
\end{equation}
It is worth noting that any spacelike two-surface $S$ determines
uniquely, up to rescalings, two null vectors orthogonal to $S$.
Any other choice of $\ell$ and $n$ will differ from the one made
in equation (\ref{eq:normals}) only by possible rescalings.  We
tie together the scalings of $\ell$ and $n$ by requiring
$\ell\cdot n = -1$.

Next, given two arbitrarily chosen spacelike orthonormal vectors
$e_1$ and $e_2$ tangent to $S$, construct a complex null vector
\begin{equation} \label{eq:mmbar}
    m :=\frac{1}{\sqrt{2}}(e_1+ie_2).
\end{equation}
It satisfies the relations $m\cdot m=0$, $m\cdot\overline{m}=1$,
$\ell\cdot m =0$, and $n\cdot m=0$.  Since $\ell$ and $n$ satisfy
$\ell\cdot n=-1$, we see that $(\ell,n,m,\overline{m})$ form a
null tetrad at $S$.  This is, of course, only one possible choice
of null tetrad, and we must ensure that physical results are
independent of this choice.
The expansions of $\ell^a$ and $n^a$ are defined as
$\theta_{(\ell)} :=q^{ab}\grad_{a}\ell_b$ and $\theta_{(n)}
:=q^{ab}\grad_an_b$, respectively.  Note that in order to find the
expansions, we only need derivatives of $\ell$ and $n$ along $S$,
and there is no need to extend the null tetrad into the full
spacetime.  However, if in some numerical computations it
is necessary to extend the null tetrad smoothly into the full
spacetime; all calculations will be insensitive to this extension.
The surface $S$ is said to be an apparent horizon if it is the
outermost outer-marginally-trapped-surface, i.e.\ it is the
outermost surface on $\Sigma$ with $\theta_{(\ell)}=0$ and
$\theta_{(n)}<0$.

Consider now the world tube of apparent horizons $H$ constructed
by stacking together the apparent horizons on different spatial
slices.  As we shall show later in section \ref{sec:trappinghor},
smooth segments of this world tube are generically spacelike; 
it is null when no matter or radiation is falling into the black
hole. At late times,
one expects the black hole to reach equilibrium when radiation and
matter are no longer crossing the horizon.  In this regime, the
world tube $H$ will be a null surface, and the two-metric $q_{ab}$
on the apparent horizon $S$ may now be viewed as a degenerate
three-metric on the null surface $H$.  Furthermore, from
experience with numerical simulations and also from general
topological censorship results (see e.g.\ \cite{fsw}), we know that
at late times, the apparent horizons must have spherical topology.
Therefore, at late times, the topology of $H$ is $S^2\times
\mathbb{R}$.  Finally, in this regime, the outward-normal $\ell$
constructed in equation (\ref{eq:normals}) is a null normal to the
world tube $H$, and most importantly, from the definition of an
apparent horizon, the outward normal $\ell$ is always expansion
free.

We will now argue that the isolated horizon framework is ideally
suited to describe apparent horizons in the regime when the world tube
$H$ is null.  For our purposes, the straightforward definition of a
\emph{non-expanding horizon} given below shall suffice.  To carry out
the Hamiltonian analysis in order to define mass and angular momentum,
we actually need to impose further conditions on non-expanding
horizons, which we shall briefly describe towards the end of this
section.  The formulae for mass and angular
momentum make sense even on non-expanding horizons.\\

\noindent\textbf{Definition}: A three dimensional sub-manifold
$\Delta$ of a space-time $(\mathcal{M},g_{ab})$ is said to be a
non-expanding horizon (NEH) if it satisfies the following
conditions:
\begin{description}
\item{(i)} $\Delta$ is topologically $S^2 \times \mathbb{R}$ and null;
\item{(ii)} The expansion $\theta_{(\ell)} := q^{ab}\nabla_a \ell_b$
of $\ell$ vanishes on $\Delta$, where $\ell$ is any null normal to
$\Delta$;
\item{(iii)} All equations of motion hold at $\Delta$ and, if any
matter field are present with $T_{ab}$ as the stress energy tensor,
then we require $-T_{b}^a\ell^b$ to be future directed and causal for
any future directed null normal $\ell^b$.
\end{description}
Note that if condition (ii) holds for one null normal $\ell$, then
it holds for all.  We will only consider those null normals which
are nowhere vanishing and future directed.  We are therefore
allowed to rescale $\ell$ by any positive-definite
function.   The energy condition in (iii) is implied
e.g.\ by the null energy condition which is commonly assumed; for the
remainder of this paper, we shall usually consider only vacuum spacetimes.

Comparing the properties of the world tube $H$ described earlier
with conditions (i) and (ii) in the definition, we see that the
NEH is precisely what we need to model the physical situation at
hand;  when the black hole is approximately isolated, the world
tube $H$ represents a non-expanding-horizon $\Delta$.  The
motivation behind the conditions in the definition are thus rather
straightforward from the perspective of apparent horizons.

We have motivated the above definition using the world tube of
apparent horizons which in general clearly depends on how we choose to foliate
the spacetime with spatial surfaces.  It is then natural to ask whether 
an NEH depends on
the choice of this foliation. To avoid unnecessary
technicalities, let us consider a possibly dynamical spacetime
containing one isolated black hole. As mentioned in the
introduction, we can choose spatial hypersurfaces on which there are
no apparent horizons so that the world tube of apparent horizons does 
not exist for this foliation.  Let us ignore this possibility and focus 
on the case when each spatial slice 
has exactly one marginally trapped surface and hence exactly one apparent
horizon; the generalization to multi black hole spacetimes is
straightforward.  It is not difficult to verify that the world 
tube of apparent horizons is
independent of the spacelike hypersurfaces used to construct it. In
other words, the null surface (the NEH) obtained by stacking 
together all the apparent horizons at different times is independent 
of how we choose to foliate the spacetime by spacelike hypersurfaces;
this is not true in the regime when the horizon is non-isolated and
the world tube is spacelike.
An NEH, even when viewed as the world tube of apparent horizons, 
is therefore an invariant notion in the four dimensional spacetime.  
Every cross section of an NEH is
potentially an apparent horizon if it arises as the intersection of
the NEH with a spatial slice.  Choosing a different spatial slice
corresponds to obtaining a different cross section of the NEH.  As we
shall see later, for calculating physical quantities such as angular
momentum and mass, it does not matter which cross section of the NEH
we choose to work with.

Even though cross sections of an NEH are very closely related to
apparent horizons, there are two differences: (i) Apparent horizons 
are required to be the \emph{outermost} surfaces on a spatial slice with the
afore-mentioned properties.  This may not be true in general for cross
sections of $\Delta$; (ii) Since they are trapped surfaces,
apparent horizons also satisfy the condition $\theta_{(n)}<0$.
Though this will most likely be true in actual numerical
simulations, it turns out that this condition is not required to
study the mechanics of isolated horizons.  In fact, there exist
exact solutions representing black holes which are isolated
horizons but do not satisfy $\theta_{(n)}<0$, e.g., the distorted
black holes studied by Geroch and Hartle \cite{gh}.  In these
solutions the integral of $\theta_{(n)}$ over a cross section of
the horizon is still negative even though $\theta_{(n)}$ is not
necessarily negative everywhere.  In the remainder of this paper,
we shall ignore these caveats and the phrases `apparent horizon'
and `cross-section of $\Delta$' will be
used interchangeably.\\

\noindent \emph{Consequences of the definition:} The simple
definition of a NEH leads to a surprisingly large number of
consequences.  The important results we shall need later are the
existence of a one-form $\omega_a$ whose definition is given in
equation (\ref{eq:defomega}), and the fact that the component
$\Psi_2$ of the Weyl tensor is a physical, i.e.\ gauge invariant,
property of an isolated horizon.  Keeping these two results in
mind, the rest of this sub-section may be skipped on a first
reading.

While stating the results, it is convenient to use a null-tetrad
$(\ell, n,m,\overline{m})$ which is \emph{adapted to the horizon};
this means that $\ell$ is a (future directed) null normal to
$\Delta$.  An example of such a null-tetrad is the one constructed
in equations (\ref{eq:normals}) and (\ref{eq:mmbar}), but there is,
of course, an infinity of such tetrads.  Physical quantities will
be independent of which null-tetrad we choose.  These consequences do
not require the spacetime to be vacuum; they only assume the energy
condition mentioned after the definition above.

\begin{itemize}
\item \emph{Area of apparent horizons is constant:} Any null normal
$\ell$ is always geodetic and is also a symmetry of the degenerate
intrinsic metric $q_{ab}$  on $\Delta$: $\Lie_\ell q_{ab}
\triangleq 0$ (the symbol `$\triangleq$' means that the
equality holds only at points of $\Delta$).  Furthermore, the area
of \emph{any} cross-section $S$ defined as
$A_{\Delta}:=\oint_S\textrm{d}^2V$ is the same where
$\textrm{d}^2V$ is the natural area measure on $S$ constructed
from the two-metric $q_{ab}$.  In particular, the area of the
apparent horizons is constant in time.
\item \emph{Definition of $\omega_a$:} It will be shown in
section \ref{sec:trappinghor} that the shear of $\ell$ defined as
$\sigma_{(\ell)}:=m^am^b\grad_a\ell_b$ is zero.  The twist of
$\ell$ is trivially zero because $\ell$ is normal to a smooth
surface, and the expansion of $\ell$ is zero by definition.  Since
the null normal $\ell$ is expansion, twist and shear free, there
must exist a one-form $\omega_a$ intrinsic to $\Delta$ such that
\begin{equation} \label{eq:defomega}
    t^a \grad_{{a}}\ell^b \triangleq t^a \omega_a\ell^b
\end{equation}
for any vector $t^a$ tangent to $\Delta$.  If the null normal
$\ell$ is rescaled $\ell\rightarrow \tilde{\ell}=f\ell$, the
one-form $\omega_a$ transforms as
\begin{equation}  \label{eq:transomega}
    \omega\rightarrow \tilde{\omega}=\omega + \textrm{d}\left( \ln
    f\right)
\end{equation}
where `$\textrm{d}$' is the exterior derivative on $\Delta$.

\item \emph{Gauge invariance of $\Psi_2$:} If we choose a null tetrad
adapted to $\Delta$ such as in equations (\ref{eq:normals}) and
(\ref{eq:mmbar}), then we get the following restrictions on the
Weyl tensor $C_{abcd}$ at $\Delta$ \cite{afk}:
\begin{equation}\label{eq:psi}
    \Psi_0 \triangleq \Psi_1 \triangleq 0
\end{equation}
and
\begin{equation}
    \textrm{d}\omega \triangleq 2\ImPt{\Psi_2}\,\epsilon,
\end{equation}
where
\begin{eqnarray}
    \Psi_{0} &=& C_{abcd}\ell^{a}m^{b}\ell^{c}m^{d}\\
    \Psi_{1} &=& C_{abcd}\ell^{a}m^{b}\ell^{c}n^{d}\\
    \Psi_{2} &=& C_{abcd}\ell^{a}m^{b}\overline{m}^{c}n^{d}
\end{eqnarray}
and $\epsilon$ is the area two-form on the apparent horizon.  This
can be used to prove the important result that $\Psi_2$ is gauge
invariant.  The gauge freedom we are concerned with here is the
choice of null tetrads adapted to $\Delta$.  The allowed gauge
transformations relating different choices are null rotations
about $\ell$:
\begin{eqnarray}\label{eq:nullrot}
    \ell &\rightarrow& \ell\nonumber\\
    m &\rightarrow& m + \overline{c}\ell\\
    n &\rightarrow& n + cm +
    \overline{c}\overline{m} + c\overline{c}\ell\nonumber
\end{eqnarray}
and spin-boost transformations:
\begin{eqnarray}\label{eq:spinboost}
    \ell &\rightarrow& A\ell\nonumber\\
    n &\rightarrow& A^{-1}n\\
    m &\rightarrow& e^{2i\theta}m\nonumber
\end{eqnarray}
Here $c$, $A$ and $\theta$ are arbitrary smooth functions on
$\Delta$.  It turns out that $\Psi_2$ is always invariant under
spin-boost transformations; under null rotations, it transforms as
(see e.g.\ the appendix of \cite{s})
\begin{equation}
    \Psi_2\rightarrow \Psi_2 + 2c\Psi_1 + c^2\Psi_0 \, .
\end{equation}
Since $\Psi_0$ and $\Psi_1$ vanish at $\Delta$, we see that
$\Psi_2$ is in fact gauge invariant at the horizon: it does not
depend on the choice of null tetrad as long as $\ell$ is one of
the null generators of $\Delta$.  This property is important
because it tells us that the value of $\Psi_2$ at the horizon does
not depend on how we choose to foliate our spacetime.  A different
spatial slice $\tilde{\Sigma}$ will lead to a different
$\tilde{\ell}$, $\tilde{n}$ and $\tilde{m}$.  The two null tetrads
will be related by a combination of the following transformations:
a null-rotation about $\ell$, a spin-boost transformation, or a
multiplication of $m$ by a phase.  Whichever null-tetrad we use to
calculate $\Psi_2$, we will get the same result.  As we shall see,
it is the imaginary part of $\Psi_2$ which is physically
interesting for our purposes.
\end{itemize}

\noindent \emph{Further Definitions:} The notion
of non-expanding horizons describes the late time behavior of
apparent horizons.  In order to define the mass
$M_\Delta$ and angular momentum $J_\Delta$ of $\Delta$, one needs
to go beyond this definition and introduce additional structures
on the horizon.  This is done via the definition of a \emph{weakly
isolated horizon} \cite{afk,abl1}.  The Hamiltonian analysis which
leads to the definitions of mass and angular momentum requires
this extra structure.  Fortunately, it turns out that the
\emph{formulae} for $M_\Delta$ and $J_\Delta$ do not depend on
this extra structure and hold true for non-expanding horizons.   We
could therefore omit these definitions entirely and simply state
the results of the calculation.  For the sake of completeness, we
shall give the basic idea behind weakly isolated horizons.  The
rest of this sub-section may be skipped without loss of
continuity.

In an NEH, the intrinsic metric $q_{ab}$ is time independent since
$\Lie_\ell q_{ab}\triangleq 0$.  There is no restriction on the time
derivatives of the extrinsic curvature of $\Delta$ or the intrinsic
connection on $\Delta$ (by `time derivative' we mean derivative along
$\ell$).  Since $\Delta$ is a null surface, there is no natural notion
of extrinsic curvature.  The closest thing to extrinsic curvature is
the tensor ${K_a}^b$ defined via $t^a{K_a}^b \triangleq
t^a\grad_a\ell^b$ for any $t^a$ tangent to $\Delta$.  This tensor is
known in the mathematics literature as the Weingarten map, and as its
definition shows, it is $\grad_a\ell^b$ with the covariant index
pulled back to $\Delta$.  From equation (\ref{eq:defomega}) we
immediately obtain ${K_a}^b\triangleq\omega_a\ell^b$.  This implies
that requiring ${K_a}^b$ to be time independent is equivalent to
requiring $\Lie_\ell\omega_a\triangleq0$.  Unfortunately, due of the
transformation property of $\omega_a$ (see equation
(\ref{eq:transomega})), it is clear that this equation is not
meaningful if all rescalings of $\ell$ are allowed.  Note that
if we restrict ourselves to rescalings that are constant on the
horizon, then $\omega_a$ is invariant.  We thus need to restrict
ourselves to an equivalence class of null normals $[\ell]$, the
members of which are related to each other by a constant, positive
non-zero rescaling.  We can then associate a unique $\omega_a$ with
$[\ell]$.  The equation $\Lie_\ell\omega_a\triangleq0$ is now
perfectly meaningful if $\ell$ is a member of $[\ell]$.  A weakly
isolated horizon is then defined as an NEH $\Delta$ equipped with such
a preferred equivalence class of null normals $[\ell]$.  Furthermore,
the $\omega_a$ associated with $[\ell]$ must be time independent:
$\Lie_\ell\omega_a\triangleq0$.  Given an NEH, we can always find such
a preferred equivalence class (but this equivalence class is not
unique).  Thus every NEH can be turned into a weakly isolated horizon.
The Hamiltonian analysis of weakly isolated horizons leads to formulae
for mass and angular momentum, and as we shall see in section
\ref{eq:manda}, these formulae are insensitive to arbitrary rescalings
of $\ell$ and thus they make sense even for non-expanding horizons.

We can introduce an even stronger definition.  A weakly isolated
horizon requires that $\omega_a$ be time independent.  It can be
seen from equation (\ref{eq:defomega}) that $\omega_a$ is also a
component of the intrinsic connection $D_a$ induced on $\Delta$ by
the four dimensional connection $\grad_a$ compatible with the
four-metric.  However, $\omega_a$ is just one component of $D_a$.
In an \emph{isolated horizon}, we require that \emph{all}
components of $D_a$ are time independent.  It turns out that
generically, this condition gives us a preferred unique
equivalence class $[\ell]$ \cite{abl2}.

\subsection{Finding non-expanding horizons}
\label{sec:trappinghor}

Our strategy to find non-expanding horizons is to locate apparent
horizons on each spatial slice and then to check whether the world
tube $H$ obtained by stacking these horizons together is a NEH.  What
remains to be checked is whether the tube is a null surface. By
definition, a surface is null if the metric $h_{ab}$ induced on this
surface has a degenerate direction, i.e.\ if there exists a vector
$X^a$ tangent to $H$ such that $h_{ab}X^b =0$; therefore, one possible
method to check whether $H$ is isolated is to construct the induced
metric $h_{ab}$ on $H$ and see if it has a zero eigenvalue.  To
construct $h_{ab}$ numerically, we have to know the two-metric
$q_{ab}$ on at least two different time slices.  Furthermore, in a
numerical simulation, $H$ will never be exactly isolated because of
numerical errors, and it is not clear how this method can quantify how
close the horizon is to being exactly isolated.  Fortunately, there is
a much simpler method which only requires the two-metric on a single
time slice and can also provide a quantitative measure of how close $H$
is to being perfectly isolated.  This method is based on the shear
$\sigma^{(\ell)}_{ab}$ of $\ell$, which is the symmetric trace-free
part of the projection of $\grad_a\ell_b$ onto the apparent horizon
$S$.  The tensor $\sigma^{(\ell)}_{ab}$ has two independent
components, and is conveniently written in terms of a single complex
number $\sigma_{(\ell)} := m^am^b\grad_a\ell_b$, where $m$ is defined
in equation (\ref{eq:mmbar}).  To calculate $\sigma_{(\ell)}$
conveniently, we simply decompose $\ell$ using equation
(\ref{eq:normals}):
\begin{eqnarray}
    \sigma_{(\ell)} &\triangleq& m^am^b\grad_a\ell_b \nonumber\\
    &\triangleq& \frac{1}{\sqrt{2}}m^am^b\grad_aT_b +
    \frac{1}{\sqrt{2}}m^am^b\grad_aR_b\, .
\end{eqnarray}
In the above equation, we have extended the notation
`$\triangleq$' to also mean that the equality
holds only at points of $H$.
The first term is just a component of the extrinsic curvature
$K_{ab}$, while the second term can be calculated on the spatial slice
by calculating the connection associated with the three-metric
$\gamma_{ab}$.  We shall now prove the following important result
concerning $\sigma_{(\ell)}$: \\

\noindent \emph{The world tube of apparent horizons is a NEH if
and only if $\sigma_{(\ell)} \triangleq0$}.\\

Readers not interested in the proof of this statement can skip
forward to equation (\ref{eq:sdef}).

To prove this statement, we need to consider the general case when
$H$ is not null.  This has been studied in great detail by Hayward
\cite{sh}.  In Hayward's terminology, the surface $H$ is
essentially a \emph{future outer trapping horizon}.  This means
that $H$ is foliated by a family of marginally trapped surfaces
(which in our case are the apparent horizons) satisfying the
relations $\theta_{{(\ell)}}\triangleq0$, $\theta_{{(n)}}<0$ and
$\Lie_n\theta_{{(\ell)}}<0$.  These are physically very reasonable
conditions, and all black holes found in simulations are expected
to satisfy them.  Incidentally, note that the condition
$\Lie_n\theta_{{(\ell)}}\triangleq 0$ can be used to define, in a quasi
local manner, what we mean by the inside and outside of the horizon. 

The proof of this statement, adapted from \cite{sh}, is then quite
simple: let $z^a$ be a vector tangent to $H$ and orthogonal to the
foliation whose leaves are the apparent horizons.  Such a vector
field may be considered to define time evolution at the horizon.
It is easy to see that, up to a rescaling, $z^a$ can be expressed
as a linear combination of $\ell^a$ and $n^a$
\begin{equation}
    z^a \triangleq \ell^a - \alpha n^a
\end{equation}
where $\alpha$ is a smooth function on $H$.  The rescaling freedom
in $z$ will be inconsequential for our purposes.  The surface $H$
is null, spacelike, or timelike if and only if $\alpha$ is zero,
positive, or negative respectively.  We will now show that
$\alpha\geq 0$.  From the definition of apparent horizons we know
that the expansion $\theta_{{(\ell)}}$ vanishes everywhere on the
horizon, therefore $\Lie_{z}\theta_{{(\ell)}} \triangleq 0$.  This
in turn gives
\begin{equation} \label{eq:alpha}
    \alpha \triangleq \frac{\Lie_{\ell}\theta_{{(\ell)}}}{\Lie_{n}\theta_{{(\ell)}}}\, .
\end{equation}
Even though $\theta_{(\ell)}$ and $\theta_{(n)}$ are so far
defined only on $H$, in equation (\ref{eq:alpha})  (and also in
the very definition of a trapping horizon) we are taking the
derivatives of these quantities along $\ell$ and $n$ which are not
necessarily tangent to $H$.  To make sense of this equation we
need to extend $\ell$ and $n$ in a neighborhood of the surface
$S$.  This can easily be done by using the unique geodesics
determined by these vectors.

The Raychaudhuri equation for $\ell$ then leads to
\begin{equation}
    \Lie_{\ell}\theta_{{(\ell)}} \triangleq -|\sigma_{{(\ell)}}|^2 - \Phi_{00}
\end{equation}
where $\Phi_{00}=\frac{1}{2}R_{ab}\ell^{a}\ell^{b}$.  If we assume
that the spacetime is vacuum, then $\Phi_{00}=0$, and therefore
$\Lie_{\ell}\theta_{{(\ell)}} \triangleq -|\sigma_{{(\ell)}}|^2$,
which along with equation (\ref{eq:alpha}) gives
\begin{equation}
    \alpha \triangleq -\frac{|\sigma_{(\ell)}|^2}{\Lie_n\theta_{(\ell)}}\, .
\end{equation}
This immediately implies that $\alpha\triangleq0$ (which is
equivalent to $H$ being null) if and only if
$\sigma_{{(\ell)}}\triangleq0$.  This is what we wanted to show.
More generally, since $\Lie_n\theta_{(\ell)}<0$, this shows that
$\alpha\geq 0$, which means that $\Delta$ is spacelike when the
shear is non-zero.

As a side remark we also show that $H$ is null if and only if the
\emph{area element} on the apparent horizons $\epsilon_{ab}$ is
preserved in time.  To show this we need the equations
\begin{equation}
    \Lie_{\ell}\epsilon_{ab} \triangleq \theta_{{(\ell)}}\epsilon_{ab}\triangleq0
    \qquad \textrm{and} \qquad \Lie_{n}\epsilon_{ab} \triangleq
    \theta_{{(n)}}\epsilon_{ab}\, .
\end{equation}
It then follows that
\begin{equation}
    \Lie_z\epsilon_{ab}  \triangleq \Lie_{\ell}\epsilon_{ab} -
    \Lie_{\alpha n}\epsilon_{ab} \triangleq -\alpha \theta_{{(n)}}\epsilon_{ab}
\end{equation}
therefore $\alpha\triangleq0$ if and only if
$\Lie_{z}\epsilon_{ab} \triangleq 0$, which is what we wanted to
prove.  This implies that if $\alpha\triangleq0$, then the area of
cross-sections of $\Delta$ is constant.  However, the converse is
not necessarily true, because $\epsilon_{ab}$ could be changing in
such a way that its integral is constant; the area can be constant
globally without being constant locally.  Thus, in principle, we
can have situations in which the area is constant without the
horizon being isolated.  However, constancy of area is still a very
useful first check to see when the horizon reaches equilibrium.
Finally, as a side remark we note that since $\alpha\geq 0$ and
$\theta_{{(n)}}<0$, we get $\Lie_{z}A_\Delta \geq 0$, which is the
area increase law.

In this paper we are interested in the case when $\sigma_{(\ell)}$
vanishes (up to numerical errors).  In order for the horizon to be
isolated we should have
\begin{equation}\label{eq:sdef}
    s := \oint_{S}|\sigma_{(\ell)}|^2\, \textrm{d}^2V = 0
\end{equation}
where $\textrm{d}^2V$ is the natural area measure on $S$
constructed from $q_{ab}$.  The quantity $s$ is dimensionless
since $\sigma_{(\ell)}$ has dimensions of inverse length.  For the
horizon to be numerically isolated, we require that $s$ converge
to zero appropriately when the numerical resolution is
increased~\footnote{We want to point out that as it stands, the
quantity $s$ can not be used as a general measure of how isolated
a given horizon is, since it is not gauge invariant.  If $s$ is not
exactly zero to begin with, a simple
rescaling of $\ell$ changes $\sigma_{(\ell)}$ and thus $s$.
However, if we are only concerned with given apparent horizons
embedded in given spatial slices, and if we agree to use equation
(\ref{eq:normals}) for defining $\ell$ thereby removing the boost
freedom, then the horizon will be close to being isolated if the
condition $s \ll 1$ is satisfied.  While this is not a
satisfactory solution for identifying the small parameter describing
a horizon which is `almost isolated', it might nevertheless be
useful in practice.  The complete solution to this problem will
require using results from \cite{ak} where the world tube of apparent
horizons is analyzed for the spacelike case.}.

\subsection{Mass and angular momentum}
\label{eq:manda}

Now let us define angular momentum.  The details and proofs of the
results below can be found in \cite{abl1}.  In order to define the
angular momentum, we need to assume that $S$ is axisymmetric,
i.e.\ there is a vector field $\varphi^a$ tangent to $S$ such that
it preserves $q_{ab}$~\footnote{This is actually an
oversimplification: in addition to preserving $q_{ab}$, the vector
field $\varphi^a$ must also preserve the additional structure
introduced to define weakly isolated horizons (discussed briefly
at the end of section \ref{sec:definitions}), namely the preferred
equivalence class $[\ell]$ and the one-form $\omega_a$ associated
with $[\ell]$ : $\Lie_\varphi\omega_a\triangleq 0$ and
$\Lie_\varphi\ell\triangleq c\ell$ where $c$ is a
positive-definite constant.  However, these additional conditions
are easy to verify in practice, and the non-trivial condition is
equation (\ref{eq:kill}).  We shall not concern ourselves with
these extra conditions in the remainder of this paper.}
\begin{equation}\label{eq:kill}
    \Lie_\varphi q_{ab}\triangleq0\, .
\end{equation}
The next section describes a simple method for finding $\varphi^a$.
Note that $\varphi^a$ is \textit{not} a Killing vector of the full
spacetime metric.  It is completely intrinsic to $\Delta$, and in fact
we only need data on the apparent horizon to find it.

Given $\varphi^a$ and $\omega_a$ (defined in equation
(\ref{eq:defomega})), the isolated horizon angular momentum $J_\Delta$ 
is defined as the surface term at $\Delta$ in the Hamiltonian which 
generates diffeomorphisms along a
rotational vector field which is equal to $\varphi^a$ at the horizon
\cite{abl2}. The expression for $J_\Delta$ is:
\begin{eqnarray}
    J_{\Delta} & = & -\frac{1}{8\pi}\oint_{S}(\omega_a\varphi^a)\,\textrm{d}^2V \nonumber\\
    & = &
    -\frac{1}{4\pi}\oint_{S}f\ImPt{\Psi_2}\,\textrm{d}^2V\label{eq:angmom},
\end{eqnarray}
where $S$ is the apparent horizon and the function $f$ is related to
$\varphi^a$ by $\partial_af=\epsilon_{ba}\varphi^b$.  In the second
equality, we have used equation (\ref{eq:psi}) and an integration by
parts.  Since $\Psi_2$ is gauge invariant, so is the angular momentum,
and in particular, it does not depend on the scaling of $\ell$ and it
thus makes sense even on a NEH.  If, in a neighborhood of $\Delta$,
there were a spacetime rotational Killing vector $\phi^a$ that
approached $\varphi^a$ at $\Delta$, then the above formula for
$J_\Delta$ would be equal to the Komar integral calculated for
$\phi^a$ \cite{abl1}.  However, the formula in equation
(\ref{eq:angmom}) is more general.  As a practical matter, even if
there were a Killing vector in the neighborhood of $\Delta$, it is
easier to use (\ref{eq:angmom}) rather than the Komar integral, since
the Komar integral requires finding the Killing vector in the full
four-dimensional spacetime.  It is also worth mentioning that, if the
vector field $\varphi^a$ is not a symmetry of $\Delta$ but is an
arbitrary vector field tangent to $S$, then $J_{\Delta}$ is still the
surface term at $\Delta$ in the Hamiltonian generating diffeomorphisms
along vector fields which are equal to $\varphi^a$ at the horizon.  But in this
case, there would be no reason to identify $J_{\Delta}$ with the
angular momentum, since conserved quantities such as mass and angular
momentum are always associated with symmetries.  However, the
existence of the axial symmetry is the least that must be true if the
horizon is to be close to Kerr in any sense.  Note that angular
momentum is a coordinate independent quantity; even if we use
corotating coordinates to describe the black hole, the black hole
still has the same angular momentum.

We now describe a form of equation (\ref{eq:angmom}) which is much
better suited for calculating $J_\Delta$ numerically.  From
equation (\ref{eq:defomega}), which is the defining equation for
$\omega_a$, we get
\begin{equation}
    t^a\omega_a \triangleq -n_bt^a\grad_{{a}}\ell^b \triangleq \ell^bt^a\grad_{{a}}n_b
\end{equation}
where, as before, $t^a$ is any vector tangent to $\Delta$.  Assume
that we have found the symmetry vector field $\varphi^a$ on the
horizon (the method we use for finding $\varphi^a$ is described
below).  From equation (\ref{eq:angmom}), we are eventually interested
in calculating $\varphi^a\omega_a$; since $\varphi^a$ is tangent to
$\Delta$, setting $t^a=\varphi^a$ we get
\begin{eqnarray}
    \varphi^a\omega_a &\triangleq& \varphi^a\ell^b\grad_an_b \nonumber\\
    & \triangleq & \frac{1}{2}\varphi^a(T^b+ R^b)\grad_a( T_b-R_b) \nonumber \\
    &\triangleq& \frac{1}{2}\varphi^a( T^b\grad_aT_b - T^b\grad_aR_b\\
    & & + R^b\grad_aT_b - R^b\grad_aR_b) \nonumber \\
    &\triangleq& \varphi^aR^b\grad_aT_b \triangleq -\varphi^aR^bK_{ab}\nonumber
\end{eqnarray}
where we have used equation (\ref{eq:normals}) along with the fact
that $N^a$ and $R^a$ are orthonormal.  In the last step, the
definition of extrinsic curvature $K_{ab}=-\gamma_a^{\
c}\gamma_b^{\ d}\grad_cT_d$ has been used where
$\gamma_{ab}=g_{ab}+T_aT_b$ is the three-metric on the spatial
slice $\Sigma$.  We have thus reduced the calculation of
$\varphi^a\omega_a$ to finding a single component of the extrinsic
curvature.  The integration of this scalar over the apparent
horizon yields the angular momentum:
\begin{equation}\label{eq:angmomk}
    J_\Delta = \frac{1}{8\pi}\oint_S (\varphi^aR^bK_{ab})\,\textrm{d}^2V \,
    .
\end{equation}
This is our final formula for the angular momentum.  This formula is
remarkably similar to the formula for the ADM angular momentum
computed at spatial infinity:
\begin{eqnarray}\label{eq:angmomadm}
    J^{\phi}_{ADM} &=&\frac{1}{8\pi}\oint_{S_\infty}(K_{ab}-
    \gamma_{ab}K)\phi^a\,\textrm{d}^2S^b\nonumber\\
    &=&\frac{1}{8\pi}\oint_{S_\infty}K_{ab}\,\phi^a\,\textrm{d}^2S^b\, .
\end{eqnarray}
The $\gamma_{ab}K$ term does not contribute, because $\phi^a$ is
tangent to $S_\infty$, which is the sphere at spatial infinity.  Since
the metric on $S_\infty$ is just the standard two-sphere metric, we
have no difficulty in choosing a $\phi^a$, and we can calculate
$J_{ADM}$ about any axis.  In contrast, since the metric on the
apparent horizon $S$ is distorted, finding $\phi^a$ is more
complicated.  Finally, as mentioned earlier, the similarity between
equations (\ref{eq:angmomk}) and (\ref{eq:angmomadm}) is not
surprising, because both quantities are defined to be surface terms in
Hamiltonians
generating diffeomorphisms along the appropriate rotational symmetry
vector fields; $J_\Delta$ is the surface term at $\Delta$ while
$J_{ADM}^\phi$ is the surface term at infinity.

Given $J_\Delta$, the horizon mass $M_\Delta$ is given by
\cite{afk,abl1}
\begin{equation}\label{eq:mass}
    M_\Delta = \frac{1}{2R_\Delta}\sqrt{R_\Delta^4 + 4J_\Delta^2}
\end{equation}
where $R_\Delta$ is the area radius of the horizon:
$R_\Delta=(A_\Delta/4\pi)^{1/2}$.  This formula depends on $R_\Delta$
and $J_\Delta$ in the same way as in the Kerr solution.  However, this
is a result of the calculation and not an assumption.  The formulae
for $J_\Delta$ and $M_\Delta$ are on the same footing as the formulae
for the ADM mass and angular momentum, because both are derived by
Hamiltonian methods.  Furthermore, under some physically reasonable
assumptions on fields near future time-like infinity ($i^+$), one can
show that $M_\Delta-M_{ADM}$ is equal to the energy radiated across
future null-infinity if the isolated horizon extends all the way to
$i^+$.  Thus, $M_\Delta$ is the mass left over after all the
gravitational radiation has left the system.  This lends further
support for identifying $M_\Delta$ with the mass of the black hole.
Further remarks regarding equation (\ref{eq:mass}) appear towards the
beginning of sec. \ref{sec:discussion}.

To summarize: in order to calculate the mass and angular momentum of
an isolated horizon, we need the following ingredients:
\begin{enumerate}
\item We must find the apparent horizon and check if the shear of
the outward null normal vanishes within numerical errors.  If it
does, then the isolated horizon formulae are applicable.

\item We need to find the quantity $K_{ab}R^b$ on $S$ and the 
symmetry vector $\varphi^a$ (assuming it exists).

\item The angular momentum is then a simple integral over the
apparent horizon given by equation (\ref{eq:angmomk}), and the mass
is a purely algebraic function given by equation (\ref{eq:mass}).

\end{enumerate}
The first and last steps are rather straightforward if we know the
location of the apparent horizon.  The only non-trivial step is the
calculation of $\varphi^a$.  In the next section, we show how it is
numerically calculated using the two-metric on $S$.

\section{Finding the Killing Vector}
\label{sec:Killingvec}

First of all, we should point out that in some numerical simulations
(especially simulations with built-in axi-symmetry) the axial symmetry
vector is already known. In that case, one can go ahead and find the
angular momentum using equation (\ref{eq:angmomk}).  However, we are
also interested in the more general case, when there is an axial
symmetry, but the coordinates used in the simulation are not adapted
to it.  In this section, we describe a general numerical method for
finding $\varphi^a$.  Our method of finding Killing vectors on the
apparent horizons is based on the Killing transport equation, which we
now describe.  This method does not depend on the fact that we are on
an apparent horizon, and it is possible that this procedure could find
Killing vectors efficiently in more general situations.  We first
describe the general method.

Let $\xi^a$ be a Killing vector on $(S,q_{ab})$, and define the
two-form $L_{ab}=\grad_{a}\xi_{b}$.  This is a two-form because of
the Killing equation $\grad_{(a}\xi_{b)}=L_{(ab)}=0$. It is then
not difficult to prove the following (see e.g.\ \cite{w})
\begin{eqnarray}
    v^a\grad_a\xi_b &=& v^aL_{ab}\nonumber\\
    \textrm{and} \qquad v^a\grad_aL_{bc} &=& {R_{cba}}^d\xi_dv^a \, .\label{eq:killtrans}
\end{eqnarray}
The reason for inserting an arbitrary vector $v^a$ will soon
become clear.  Instead of viewing these as equations for a Killing
vector, let us instead think of them as equations for an arbitrary
vector $\xi^a$ (or a one-form $\xi_a$) and an arbitrary two-form
$L_{ab}$. If we start with a one-form $\xi_a^{(p)}$ and a two-form
$L^{(p)}_{ab}$ at a point $p$ on the manifold, then the above
equations can be solved along any curve $\gamma(t)$ (with $v^a$ as
its tangent) starting at $p$ to give a unique one-form $\eta_a$
and a unique two-form $\alpha_{ab}$ at any other point on the
curve.  This procedure is analogous to parallel transport, but the
differential equation used in the transport is not the geodesic
equation, but instead equation (\ref{eq:killtrans}) above, and
instead of transporting a vector, these equations transport a
one-form and a two-form.  Viewed this way, equations
(\ref{eq:killtrans}) are often referred to as the \emph{Killing
transport equations} \cite{am}.  We are thus led to consider the
vector space $V_p$ consisting of all pairs
$(\xi^{(p)}_a,L^{(p)}_{ab})$ for an arbitrary one-form
$\xi^{(p)}_a$ and an arbitrary two-form $L^{(p)}_{ab}$ at a point
$p$.  For any curve $\gamma(t)$ which starts at $p$ and ends at
$q$, the equations in (\ref{eq:killtrans}), being linear, give us
a linear mapping between $V_p$ and $V_q$.  If
$(\xi^{(p)}_a,L^{(p)}_{ab})\in V_p$ actually comes from a Killing
vector and its derivative, then it will be mapped to
$(\xi^{(q)}_a,L^{(q)}_{ab})\in V_q$, which comes from the same
Killing vector.

If we consider closed curves starting and ending at the point $p$,
then the Killing transport for a curve $\gamma(t)$ gives us a linear
mapping $M_p(\gamma):V_p\rightarrow V_p$.  A Killing vector
corresponds to an eigenvector of $M_p(\gamma)$ with eigenvalue equal
to unity for any closed curve $\gamma$.  Note that an eigenvector $Z$ of
$M_p(\gamma)$ with unit eigenvalue need not necessarily correspond to a
Killing vector; this will be true only if $Z$ is an eigenvector of
$M_p(\gamma)$ with unit eigenvalue for \emph{every} sufficiently smooth
curve $\gamma$ beginning and ending at $p$.  However, if there exists
only one eigenvector with eigenvalue $1$, it is the only possible
candidate for a Killing vector and it is easy to check explicitly (by
calculating the appropriate Lie derivative of $q_{ab}$) whether this
candidate is indeed a Killing vector.  If there exists no such
eigenvector, then it tells us that $q_{ab}$ does not have any Killing
vectors at $p$.

In our case, $S$ is a
topological two-sphere which means that $V_p$ is a three dimensional
vector space and, if we choose a basis, $M_p(\gamma)$ can be
represented as a $3\times 3$ matrix (if $S$ is an $n$ dimensional
manifold, then $M_p(\gamma)$ is a $\frac{1}{2}n(n+1)$ dimensional
matrix).  Finding the Killing vector at $p$ then reduces to an
eigenvalue problem for a $3\times 3$ matrix.  For a constant curvature
two-sphere, as in the Schwarzschild horizon, this matrix will just be
the identity matrix for any point $p$.  For an axially symmetric
sphere, such as the one in a Kerr spacetime, there will be precisely
one such eigenvector.  Having found $\xi^a$ and $L_{ab}$ at one point,
we can again use equation (\ref{eq:killtrans}) to find it everywhere
on the sphere, using various other curves.  Finally, the Killing vector
is normalized by requiring its integral curves to have affine length
$2\pi$ (it can be shown that the integral curves must in fact be
closed).  Since we are only free to rescale the Killing vector by an
overall constant, we only have to perform the normalization on one
integral curve.  This normalization is valid for rotational Killing
vectors.  If we were dealing with, say, translational or stationary
Killing vectors, the appropriate normalization condition would be to
require the vector to have unit norm at infinity.

To make this procedure concrete, let us write down the equations
explicitly in spherical coordinates.  Let $(\theta,\phi)$ be arbitrary 
spherical coordinates on $S$ with $\theta\in [0,\pi]$ being the
azimuthal angle and $\phi\in [0,2\pi]$ being the polar angle. 
In these coordinates, the Riemannian two-metric
$q_{ab}$ on $S$ is:
\begin{equation}
    q=q_{\theta\theta}\,d\theta\otimes d\theta +
    q_{\phi\phi}\,d\phi\otimes d\phi + q_{\theta\phi}\,(d\theta\otimes
    d\phi+ d\phi \otimes d\theta) \, .
\end{equation}
The horizon may be arbitrarily distorted; $q_{ab}$ does not have
to be the standard two-sphere metric.  Note that on a sphere, any
two-form $L_{ab}$ can be written uniquely as
$L_{ab}=L\epsilon_{ab}$, where $L$ is a function on $S$, and
$\epsilon_{ab}$ is the area two-form on $S$; $\epsilon =
\sqrt{\det q}\,d\theta\wedge d\phi$ where ${\det q =
q_{\theta\theta}q_{\phi\phi} - q_{\theta\phi}^2 }$ is the
determinant of $q_{ab}$.  Any one-form $\xi_a$ can be expanded as
$\xi = \xi_{\theta}d\theta + \xi_{\phi}d\phi$.  The covariant
derivative of a one-form $\xi_a$ is expressed in terms of the
Christoffel symbols ${\Gamma_{ab}}^c$ as
\begin{equation}
    \grad_a\xi_b=\partial_a\xi_b -{\Gamma_{ab}}^c\xi_c \, .
\end{equation}
The Riemann tensor of $q_{ab}$ has only one independent component
\begin{equation}
    R_{abcd}=\frac{1}{2}R\epsilon_{ab}\epsilon_{cd}\, .
\end{equation}
Now we must choose a closed curve in order to find the Killing
vector at a single point.  The equator $(\theta=\pi/2)$ is a
convenient choice for the curve since it avoids the coordinate
singularity at the poles.  The tangent vector $v^a$ is then simply
$\partial_{\phi}$.  The equations (\ref{eq:killtrans}) then become
\begin{eqnarray} \label{eq:phitrans}
    \frac{\partial \xi_\theta}{\partial\phi}
    &=& {\Gamma_{\theta\phi}}^\theta\xi_\theta +
    {\Gamma_{\theta\phi}}^\phi\xi_\phi - L\sqrt{\det q} \, ,\nonumber \\
    \frac{\partial \xi_\phi}{\partial\phi}&=&
    {\Gamma_{\phi\phi}}^\theta\xi_\theta +
    {\Gamma_{\phi\phi}}^\phi\xi_\phi \, ,\nonumber  \\
    \frac{\partial L}{\partial \phi} &=&
    \frac{1}{2}R\sqrt{\det q}\left(q^{\theta\theta}\xi_\theta +
    q^{\theta\phi}\xi_\phi\right) \, .
\end{eqnarray}
(In these formulas we do not sum over repeated indices.)  These are
three coupled, linear, first-order differential equations in
$(\xi_\theta,\xi_\phi,L)$.  The same equation holds for any line of
latitude ($\theta=$ constant).  The second-order Runge-Kutta method
was used to solve these equations.  The initial data required for
this equation are the values of $(\xi_\theta,\xi_\phi,L)$ at, say,
$\phi=0$.  The solution of the equation will be
$(\xi_\theta,\xi_\phi,L)$ at $\phi=2\pi$.  We are eventually
interested only in $(\xi_\theta,\xi_\phi)$, but the function $L$ is
necessary to transport the data.  The solution to these equations
can be written in terms of a matrix $\mathbf{M}$:
\begin{equation} \label{eq:matrix}
    \left(\begin{array}{c} \xi_{\theta} \\
    \xi_{\phi}\\ L \end{array}\right)_{(\phi=2\pi)} = \mathbf{M}
    \left(\begin{array}{c} \xi_{\theta} \\
    \xi_{\phi}\\ L \end{array}\right)_{(\phi=0)}\, .
\end{equation}
To find the matrix $\mathbf{M}$, we start with the initial data
sets $(1,0,0)$, $(0,1,0)$ and $(0,0,1)$.  The solutions will give
the first, second and third columns respectively of $\mathbf{M}$.
Next we find the eigenvalues and eigenvectors of $\mathbf{M}$.
The eigenvector with unit eigenvalue is what we want ~\footnote{In
principle, we should verify that \emph{every} closed curve
starting and ending at the point $(\theta=\pi/2,\phi=0)$ gives the
same eigenvector.  However, as in any numerical method, we only do
this for a small number of curves.  The eigenvector obtained in
this way is the only possible candidate for a Killing vector.}.
Numerically, no eigenvalue is exactly equal to unity; therefore,
in practice, we choose the eigenvalue closest to unity (see the
next section).  For the horizon to be axisymmetric or close to
Kerr in any sense, this eigenvalue should be very close to unity.
If this is not the case, then this proves that the horizon is not
close to Kerr in any sense.  All eigenvalues will be unity in the
spherically symmetric case.

Having found the eigenvector at the point $\phi=0$, we then
transport it to every grid point on the sphere.  The curves used to
transport the eigenvector are the lines of latitude and longitude.
Transport along constant $\theta$ curves is done by equations
(\ref{eq:phitrans}), while for the constant $\phi$ curves we use:
\begin{eqnarray} \label{eq:thetatrans}
     \frac{\partial \xi_\theta}{\partial \theta} &=&
    {\Gamma_{\theta\theta}}^\theta\xi_\theta +
    {\Gamma_{\theta\theta}}^\phi\xi_\phi \, ,\nonumber \\
    \frac{\partial \xi_\phi}{\partial\theta} &=&
    {\Gamma_{\theta\phi}}^\theta\xi_\theta +
    {\Gamma_{\theta\phi}}^\phi\xi_\phi + L\sqrt{\det q}
    \, , \nonumber \\
    \frac{\partial L}{\partial \theta} &=&
    -\frac{1}{2}R\sqrt{\det q}\left(q^{\phi\theta}\xi_\theta +
    q^{\phi\phi}\xi_\phi \right) \, .
\end{eqnarray}
(Again, no summation over repeated indices.)  Finally, having found
$(\xi_\theta,\xi_\phi,L)$ at each grid point, we now need to
normalize the Killing vector $\xi=(\xi_\theta,\xi_\phi)$ so that
its integral curves have affine length $2\pi$.  To do this, we
need to follow the integral curves of $\xi^a$:
\begin{equation} \label{eq:normalize}
    \frac{d\theta}{dt} = \xi^\theta(\theta,\phi) \qquad
    \textrm{and} \qquad \frac{d\phi}{dt} = \xi^\phi(\theta,\phi)
\end{equation}
and normalize the affine parameter $t$ so that its range is
$[0,2\pi]$.  Numerically, we only have to make sure that $\xi^a$ does
not vanish at the starting point.  While solving equation
(\ref{eq:normalize}) numerically, we will need the value of $\xi^a$ at
points not included in the grid.  We use a second order interpolation
method for this purpose.  This finally gives us the normalized
symmetry vector $\varphi^a$, which is used to calculate $J_\Delta$ from
equation (\ref{eq:angmomk}).

\section{Computation of $J_\Delta$ and tests of the Numerical Code}
\label{sec:numeric}

In this section, we apply our approach to finding the Killing
vector and our ability to identify an isolated horizon.
In order to validate our approach for identifying Killing vectors on
$S$, we first test our method using analytic data
($\gamma_{ij},K_{ij}$) in a simple case for which the location of $S$
and its Killing vectors are also known; we consider the boosted
Kerr-Schild solution \cite{HuMash} with the basic parameters of mass
$M=1$, spin $a=1/2$, and a boost in the $z$-direction.  The four-metric has
the form $g_{ab}=\eta_{ab} + 2H\ell_a\ell_b$ where $\eta_{ab}$ is a flat
Lorentzian metric, $H$ is a smooth function and $\ell_a$ is null
(see e.g.\ \cite{chandra} for a detailed discussion
of such metrics).  The notion of
a boost is well defined for metrics in the Kerr-Schild form because of
the presence of the flat background metric $\eta_{ab}$.  By boosting
the black hole,
we impose a coordinate distortion on the horizon, while retaining its
physical properties.  For these test cases, we know
that the horizon is isolated; and we take advantage of only
needing to compute quantities intrinsic to a two-sphere and use
spherical coordinates.  The following steps were used to test the
numerical code maintaining second order accuracy at each step:
\begin{enumerate}
\item From the Kerr-Schild data, calculate analytically the apparent
horizon two-metric $q_{ab}$, the normal $R_a$, and the components of
the extrinsic curvature $K_{ab}$ at the location of the apparent
horizon.  Discretize these quantities using a spherical grid on the
apparent horizon.
\item Using the discretized data, find the unnormalized Killing
vector, $\xi^a$, at a single point
(in our case we choose this point to be $(\theta=\pi/2,\phi=0)$)
through the procedure described in the previous section, applying
the Runge-Kutta method.
\item Solve both equations (\ref{eq:phitrans}) and (\ref{eq:thetatrans})
to find $\xi^a$ everywhere on the apparent horizon.
\item Normalize the Killing vector, $\phi^a$, using
interpolation and a Runge-Kutta method for equations
(\ref{eq:normalize}).
\item Calculate $J_\Delta$  via equation (\ref{eq:angmomk}), using $R^a$
given by the apparent horizon and $\phi^a$ determined by steps 1--4.
\end{enumerate}

\begin{figure}
    \begin{center}
    \includegraphics[height=6cm]{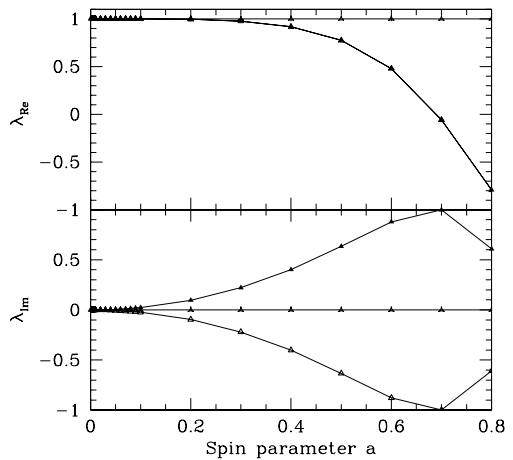}
    \caption{\small{Plots of the real and imaginary parts of the eigenvalues of
    the matrix $\mathbf{M}$ (defined in eqn. (\ref{eq:matrix}))
    versus a large range of the spin parameter $a$.}}\label{fig:lambda1}
    \end{center}
\end{figure}

\begin{figure}
    \begin{center}
    \includegraphics[height=6cm]{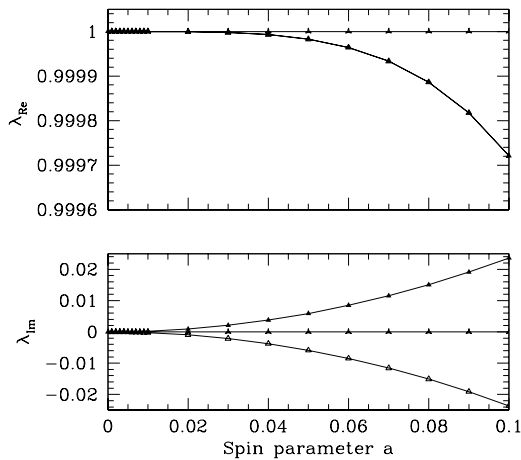}
    \caption{\small{Plots of the real and imaginary parts of the eigenvalues of the
    matrix $\mathbf{M}$
    versus the spin parameter $a$, when $a$ is small.}}\label{fig:lambda2}
    \end{center}
\end{figure}

The first step is easy if we have the analytic expressions for the
relevant quantities.  In the second step, we have to find the
matrix $\mathbf{M}$ described in equation (\ref{eq:matrix}) and
find its eigenvector with eigenvalue closest to unity.  One can
ask whether there is any ambiguity in choosing the right
eigenvalue; is it possible that more than one eigenvalue is close
to unity?  In the spherically symmetric case ($a=0$), all
eigenvalues are equal to unity and it is immaterial which one we
choose; the angular momentum will be zero.  When $a$ is
sufficiently large, one eigenvalue is much closer to unity in
magnitude as compared to the other two. In our case, it turns out
that the matrix $\mathbf{M}$ has one real eigenvalue $\lambda$ and
two complex eigenvalues $\lambda_{Re} \pm i\lambda_{Im}$.
Figure~\ref{fig:lambda1}, a plot of the real and imaginary parts
of the eigenvalues as a function of $a$ (for the un-boosted
Kerr-Schild hole) , demonstrates the unambiguous nature of the
eigenvalue for large values of $a$. The ambiguity may arise when
$a$ is very small. In figure~\ref{fig:lambda2}, we plot both
functions for a smaller range of $a$. Both plots were generated
for a resolution of $d\phi=\pi/80$.  The figures show that the
correct eigenvalue is typically easy to identify, because the
other eigenvalues diverge from unity rather rapidly and also, at
least in this case, the `wrong' eigenvalues are complex while the
correct eigenvalue is real.

\begin{figure}
    \begin{center}
    \includegraphics[height=6cm]{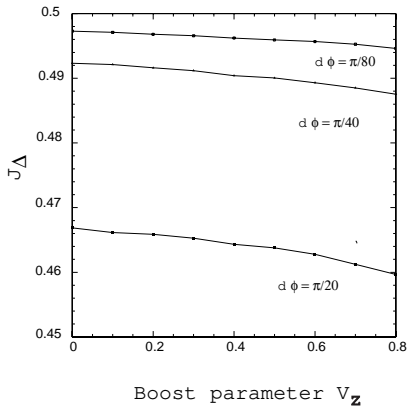}
    \caption{\small{The numerically computed angular momentum of the
    black hole at different
    boosts for a black hole with mass $M=1$ and spin $a=1/2$.  Three
    different resolutions $d\phi$ are shown.}}\label{fig:boost}
    \end{center}
\end{figure}

Having found the correct eigenvector and therefore the Killing vector
at a single point, we then find it at every other grid point and use
it to calculate $J_\Delta$.  Figure \ref{fig:boost} plots the values
of the angular momentum of the black hole found using
equation (\ref{eq:angmomk}) versus different values of the boost parameter
for the Kerr-Schild data.  Three different resolutions are plotted,
showing a second-order convergence rate towards the
known analytical value of $J_\Delta=0.5$ as expected.  Although there
is a slight
loss in accuracy as the boost approaches the speed of light, the
angular momentum loses only $1\%$ in accuracy for the least resolved
case ($d\phi=\pi/20$) in figure \ref{fig:boost} when the boost parameter is 
increased from $0$ to $0.8$.  This loss in accuracy is purely due to
numerical effects and is smaller for the higher
resolution cases.  We also obtained similar results for boosts in other
directions.

\begin{figure}
    \begin{center}
    \includegraphics[height=6cm]{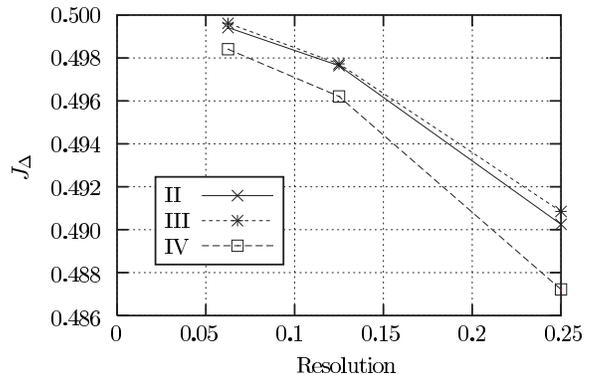}
    \caption{\small{Resolution tests for the angular momentum $J_\Delta$
    of the horizon.  The scenarios II -- IV are explained in table
    \ref{tab:table3}.}}\label{fig:angmom}
    \end{center}
\end{figure}

\begin{figure}
    \begin{center}
    \includegraphics[height=6cm]{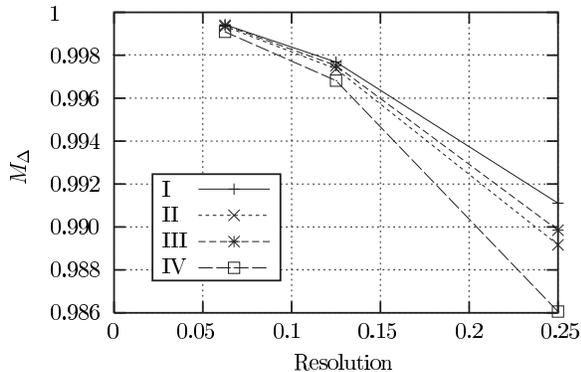}
    \caption{\small{Resolution test for the mass $M_\Delta$ of the
    horizon.  The scenarios II -- IV are explained in table
    \ref{tab:table3}.}}\label{fig:masses}
    \end{center}
\end{figure}

\begin{figure}
  \begin{center}
  \includegraphics[height=6cm]{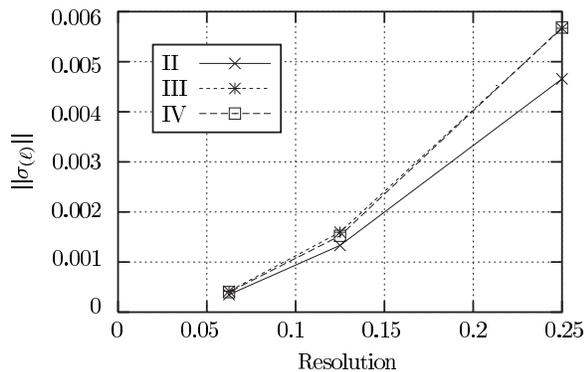}
  \caption{\small{This graph shows the $L_2$ norm of $\sigma_{(\ell)}$.
  We see that it converges to zero, indicating that the horizon is
  isolated.  The scenarios II -- IV are explained in table
    \ref{tab:table3}.}}\label{fig:sigma}
  \end{center}
\end{figure}

A more realistic situation is to compute $J_\Delta$ and $M_\Delta$
during a numerical simulation of a black-hole spacetime, in which the
spacetime data will be given on a spatial grid.  To test our method in
this case, we again use boosted Kerr-Schild data, but this time we
start with numerical data discretized on a Cartesian mesh on a spatial
slice; this mesh will not coincide with the spherical mesh on the
apparent horizon.  We use an apparent horizon finder to locate the
apparent horizon $S$ and its unit spacelike normal $R^a$, and
construct a spherical grid on the apparent horizon.  Let $dx$ and
$d\phi$ be the grid spacing of the Cartesian and spherical grid
respectively.  We want the two grids to be of similar spacing, i.e.\
we choose $d\phi$ such that $d\phi\approx dx/R$ where $R$ is the
coordinate radius of the apparent horizon.  The data are then
interpolated onto the spherical grid, an additional source of error.
We extract the two-metric
$q_{ab}$ numerically from the data, use it to find the Killing vector
field $\varphi^a$, and apply our formula for angular momentum.  We
present a series of test cases involving the black hole in boosted
Kerr-Schild data.  One is static, and three others have a spin of
$1/2$ about the $z$-axis, with one of the spinning holes boosted
perpendicular to the spin along the $x$-axis, another parallel to the
spin along the $z$-axis.  Table \ref{tab:table3} lists the different
scenarios.
\begin{table}
\caption{\label{tab:table3} Various scenarios}
\begin{ruledtabular}
\begin{tabular}{lllll}
Scenario & $M$   & $a$ & $v_x$ & $v_z$ \\\hline
I   & 1 & 0 & 0 & 0 \\
II  & 1 & 1/2   & 0 & 0 \\
III & 1 & 1/2   & 1/2   & 0     \\
IV  & 1 & 1/2   & 0 & 1/2
\end{tabular}
\end{ruledtabular}
\end{table}
Due to the additional complexity of having the data in a mesh that is
not the one on $S$ where the calculation is done, we have to deal with
two different numerical grids.  We refined both the Cartesian grid and
the spherical grid intrinsic to the apparent horizon to perform
convergence tests.  All runs were performed with three resolutions:
1. $dx=1/4$, $d\phi=10^\circ$; 2. $dx=1/8$, $d\phi=5^\circ$; 3.
$dx=1/16$, $d\phi=2.5^\circ$.  All lengths are measured in units of
mass $M$ where $M$ is the mass of the black hole; $J_\Delta$ is
measured in units of $M^2$.  Figure \ref{fig:angmom} shows $J_\Delta$
versus resolution, and figure \ref{fig:masses} displays $M_\Delta$ versus
resolution, showing second-order convergence to the known solutions for
each of the cases described in table \ref{tab:table3}.
In addition to $J_\Delta$ and $M_\Delta$,  we also monitor how
well we converge to a truly isolated horizon, one in which the
shear $\sigma$ is zero.  Figure \ref{fig:sigma} plots the value
of $\sigma$ versus resolution and demonstrates second-order
convergence toward zero.  As expected due to additional errors in the
apparent horizon tracker, the convergence factors are not as
good as in the case of analytic data; but are still acceptable for
second order convergence.

\section{Discussion}
\label{sec:discussion}

In this section we want to compare our method of finding the mass
and angular momentum of a black hole in a numerical simulation
with other methods that are commonly used.

Note that the method proposed in this paper has three advantages:
(i) it is not tied to a particular geometry (like the Kerr
geometry), (ii) it is completely coordinate independent, and (iii)
it only requires data that is intrinsic to the apparent horizon.
The commonly used alternatives for calculating mass and angular
momentum do not share all three of these features.

The reader might object to (i) on the grounds that our
formula for $M_\Delta$ (equation (\ref{eq:mass})) is also motivated by 
the corresponding Kerr formula.  In this regard, the following remarks
are worth mentioning. Even though our formula for 
$M_\Delta$  is based on the corresponding
Kerr equation, much less use is made of the specific properties of a
Kerr black hole
than in most conventional methods such as the great circle method
described below.  In particular, while deriving equation (\ref{eq:mass}),
we do not make any assumption regarding the near horizon geometry.  
All that is required is that the members of the 
Kerr family are elements of the phase space under consideration (see
\cite{afk,abl1} for a detailed specification of the phase space).  The
Kerr solutions are used only to provide a reference point for values of the
black hole
energy; this is somewhat analogous to choosing the zero of a potential
in classical particle mechanics.  In principle, we could also have
chosen a different two parameter family of solutions but we choose the
Kerr family due to its well known importance in black hole theory.

Owing to the uniqueness theorems of classical general relativity,
it is commonly believed that a black hole that has been created in
a violent event will radiate away all its higher multipole moments
and settle down to form a Kerr black hole near the horizon.  One
strategy for assigning a mass and an angular momentum to a black
hole is then to identify the member of the Kerr family one is
dealing with and to read off the corresponding mass and angular
momentum parameters.

While this strategy is physically well motivated, and one does
expect the final black hole to be close to Kerr in some sense, we
can refine this strategy considerably.  The main difficulty with
this method is that there are many subtleties and open questions
regarding the issue of uniqueness of the final black hole.  To
briefly illustrate this point, let us consider the isolated
horizon describing the final black hole.  The intrinsic geometry
of the horizon is, by definition, time independent.  However, it is
not necessary that \emph{four dimensional} quantities evaluated at
the horizon must also be time independent.  For instance, using
the Einstein equations at the isolated horizon, it turns out that
the expansion and shear of the inward pointing normal $n^a$ (see
equation (\ref{eq:normals})) may not be time independent; these
quantities decay exponentially \cite{abl2}.  This means that the
four-geometry in the vicinity of a NEH is generically not
time independent, and hence may not be isometric to a Kerr
solution.  It is also not clear whether the four geometry tends to
the Kerr geometry as we approach future time like infinity.
Clearly, we need to prove a statement which says:  `sufficiently far'
into the future, the spacetime metric in a `sufficiently small'
neighborhood of the horizon, is `close to the Kerr metric'.  In this
statement, each phrase within quotes is ill defined.
To prove or disprove such statements, it is imperative, at the very least,
that one does not start with the assumption that the final black hole
is Kerr.  In
particular, it is desirable that we not use geometrical properties of
the Kerr horizon to calculate angular momentum and mass.

If one nevertheless makes the assumption that the final black hole
is described by a Kerr geometry, one has to find a way to identify
the particular member in the Kerr family.  The method that is most
commonly used is the \emph{great circle method} which is based on
properties of the Kerr horizon found by Smarr \cite{smarr}.  It can
be described as follows:

In the usual coordinates, let $L_e$ be the proper length of the
equator of the Kerr apparent horizon
and $L_p$ the proper length of a polar meridian.  Here the equator is the
great circle of maximum length and a polar meridian is a great
circle of minimum length.  The distortion parameter $\delta$ is
then defined to be $(L_e-L_p)/L_e$.  Smarr then showed that the
knowledge of $\delta$, together with one other quantity like area,
$L_e$, or $L_p$, is sufficient to find the parameters $m$ and $a$
of the Kerr geometry.

The difficulty with this method, apart from relying overly on
properties of the Kerr spacetime, is that notions such as great
circles, equator or polar meridian etc.\ are all highly coordinate
dependent.  If we represent the familiar two-metric on the Kerr
horizon in different coordinates, the great circles in one
coordinate system will not agree with great circles in the other
system.  The two coordinate systems will therefore give different
answers for $M$ and $a$ as calculated by this method.  In certain
specific situations where one has a good intuition about the
coordinate system being used and the physical situation being
modelled, this method might be useful as a quick way of
calculating angular momentum, but it is inadequate as a
general method.

The problem of coordinate dependence can be dealt with in
axisymmetric situations without assuming that the coordinate system used in
the numerical code is adapted to the axial symmetry.  The idea
is to use the orbits of the Killing vector as analogs of the lines
of latitude on a metric two-sphere.  The analog of the equator is
then the orbit of the Killing vector which has maximum proper
length.  This defines $L_e$ in an invariant way.  The north and
south poles are the points where the Killing vector vanishes, and
the analog of $L_p$ is the length of a geodesic joining these two
points (because of axial symmetry, all geodesics joining the poles
will have the same length).  Since this geodesic is necessarily
perpendicular to the Killing vector, we just need to find the
length of a curve which joins the north and south poles and is
everywhere perpendicular to the Killing orbits.  With $L_e$ and
$L_p$ defined in this coordinate invariant way, we can follow the
same procedure as in the great circle method to calculate the mass
and angular momentum.  This method can be called the
\emph{generalized} great circle method.

How does the generalized great circle method compare to our
method?  From a purely practical point of view, note that this
method requires us to find the Killing vector, to determine the
orbit of the Killing vector with maximum length, and to calculate
the length of a curve joining the poles which is orthogonal to the
Killing orbits.  The first step is the same as in the isolated
horizon method presented in this paper.  While the next step in the
isolated horizon method is simply to integrate a component of the
extrinsic curvature on the horizon, this method requires more work,
and furthermore, the numerical errors involved are at least as
high as in the isolated horizon method.  Thus the simplicity of
the great circle method is lost when we try to make it coordinate
invariant, and it retains the disadvantage of relying heavily on
the properties of the Kerr geometry.

It should also be mentioned here that there exist exact solutions
to Einstein's equations representing static, non-rotating and
axisymmetric black holes.  Examples of such solutions are the
distorted black hole solutions found by Geroch and Hartle
\cite{gh} or the solutions representing black holes immersed in
electromagnetic fields \cite{ernst}.  The apparent horizons in all
these solutions are distorted, and the generalized great circle
method will give a non-zero value for the angular momentum. Our method
based on equations (\ref{eq:angmomk}) and (\ref{eq:angmom}) will 
give $J_\Delta =0$
because the imaginary part of $\Psi_2$ vanishes at the horizon for
these solutions.  While
these solutions are probably not relevant for numerical simulations of
binary black hole collisions, they represent physically
interesting situations in which a black hole is surrounded by
different kinds of external matter fields which distort the black
hole; these black holes may have some relevance astrophysically.
These solutions show that the generalized great circle method
cannot be correct in general.  They also illustrate that the great
circle method will in general give results that are different from
the ones obtained using Komar integrals.

A completely different approach to finding the mass and angular
momentum of a black hole in a numerical solution is to use the
concept of a Killing horizon.  Since in a numerical simulation one
is interested in highly dynamical situations, one can not assume
the existence of Killing vectors in the whole spacetime.  Instead
one assumes that stationary and axial Killing vectors exist in a
neighborhood of the horizon, and then uses appropriate Komar
integrals to find the mass and angular momentum.

While this method is coordinate independent and does not rely on a
specific metric, it has two disadvantages when compared with the
isolated horizon approach.  First, it is not a priori clear how the
stationary Killing vector is to be normalized if it is only known
in a neighborhood of the horizon.  Secondly, this method requires
the Killing vectors to be known in a whole neighborhood of the
horizon.  Computationally this is more expensive than finding a
Killing vector just on the horizon.  Conceptually it is also
unclear how big this neighborhood of the horizon should be.
Furthermore, at present there is no Hamiltonian framework
available in which the boundary condition involves the existence
of Killing vectors in a finite neighborhood of the horizon.  In a
sense, the isolated horizon framework extracts just the minimum
amount of information from a Killing horizon in order to carry out
the Hamiltonian analysis and define conserved quantities.

\section{Applications and Future Directions}
\label{sec:conclusion}

One situation where the calculation of $M_\Delta$ and $J_\Delta$ might
be useful is in studying properties of initial data.  Consider, for
example, an initial data representing a binary black hole system.  If
the black holes are far apart, then we may consider them to be
isolated (this can be verified, for example, by calculating the shear
of the inward null normal as in equation (\ref{eq:sdef})).  Then the
difference $E=M_\Delta^1 + M_\Delta^2 - M_{ADM}$ represents the
binding energy between the holes plus the energy due to radiation and
possibly also the kinetic energy of the black holes.  In
order to find the individual black hole masses $M_\Delta^1$ and
$M_\Delta^2$ equations (\ref{eq:angmomk}) and (\ref{eq:mass}) will be
useful for this purpose.  It is also interesting to compare different
initial data sets which represent roughly the same physical situation
by calculating the quantity $E$.  A lot of work in this direction has
recently been carried out by Pfeiffer et al.\ \cite{pct}.  The isolated
horizon framework may provide some further insights.

The isolated horizon framework may also be used to construct
initial data representing two (or more) black holes far away from
each other.  We want to specify the individual black hole spins,
velocities and masses in the initial data when the two black holes
are very far apart.  For this purpose, the formulae for $J_\Delta$
and $M_\Delta$ would be relevant, and we may assume that the black
holes will be isolated at least for a short time and will form an
isolated horizon.  The isolated horizon conditions will then yield
boundary conditions at the apparent horizon which can be used to
solve the constraint equations on the spatial slice.  In fact,
pioneering work in this direction has already been carried out by
Cook \cite{c}; the quasi-equilibrium boundary conditions developed
by Cook are identical to the isolated horizon boundary conditions
in many ways.

Let us also briefly discuss one more important future application:
extracting radiation waveforms.  We expect that information about
gravitational radiation will be encoded in a component of the Weyl
tensor such as $\Psi_4
=C_{abcd}n^a\overline{m}^bn^c\overline{m}^d$ where $n^a$ and $m^a$
are members of a null tetrad.  In certain situations where we have
a background metric or if we have Killing vectors, there may be a
natural choice of the null tetrad used to calculate $\Psi_4$.
However, in general situations, it is not clear how to construct
this preferred null tetrad.  Without a well defined way of
calculating radiation waveforms, it is very hard to even compare
the results of different simulations which model similar physical
situations but in different coordinates.  It turns out that the
isolated horizon framework could be used to provide a preferred
coordinate system in the vicinity of an isolated horizon.  For
this purpose, we need more structure on the horizon than provided
by a NEH; we need the notion of an \emph{isolated horizon} as
discussed briefly at the end of section \ref{sec:definitions}.  It
can be shown that, generically, an isolated horizon has a preferred
foliation which may be said to define its rest frame.  In general,
this preferred foliation will not agree with the foliation given
by the apparent horizons in a particular choice of Cauchy surfaces
in spacetime.  However, this preferred foliation can be constructed
in a completely coordinate independent manner.  Given this
preferred foliation, we can construct a coordinate system
analogous to the Bondi coordinates constructed near null infinity
\cite{letter,abl2}.  This preferred coordinate provides an
invariant way of calculating radiation waveforms and comparing the
results of different simulations.

\section{Acknowledgements}

We would like to thank Abhay Ashtekar and Pablo Laguna for
countless discussions, criticisms and suggestions.  We are also
grateful to Doug Arnold, Greg Cook, Harald Pfeiffer, and Jorge
Pullin for useful discussions and suggestions.  This work was
supported in part by the NSF grants PHY01-14375, PHY-0090091 and
PHY-9800973  and by the DFG grant SFB-382.  BK was also supported
by the Duncan Fellowship at Penn State.  All the authors
acknowledge the support of the Center for Gravitational Wave
Physics, which is funded by the National Science Foundation under
Cooperative Agreement PHY-0114375.

\end{document}